\documentclass[%
 aip,
 cha,
 amsmath,amssymb,
 reprint,%
]{revtex4-1}

\usepackage{graphicx}
\usepackage{dcolumn}
\usepackage{bm}

\usepackage[utf8]{inputenc}
\usepackage[T1]{fontenc}
\usepackage{mathptmx}
\usepackage{etoolbox}

\usepackage{diagbox}

\usepackage{amssymb, amsmath, amsfonts}
\usepackage{mathtools}
\usepackage{color}
\usepackage{calrsfs}
\DeclareMathAlphabet{\pazocal}{OMS}{zplm}{m}{n}

\usepackage{xspace, units}

\usepackage{lineno}
\usepackage{float}
\usepackage{booktabs}
\usepackage{multirow}

\newcommand{\defi}{\mathrel{\mathop:}=}

\newcommand {\CEV}{\ensuremath{\pazocal{C}^{\mathrm{E}}_{\mathrm{v}}}\xspace}
\newcommand {\CEE}{\ensuremath{\pazocal{C}^{\mathrm{E}}_{\mathrm{e}}}\xspace}
\newcommand {\CEVE}{\ensuremath{\pazocal{C}^{\mathrm{E}}_{\mathrm{v,e}}}\xspace}
\newcommand {\CBV}{\ensuremath{\pazocal{C}^{\mathrm{B}}_{\mathrm{v}}}\xspace}
\newcommand {\CBE}{\ensuremath{\pazocal{C}^{\mathrm{B}}_{\mathrm{e}}}\xspace}
\newcommand {\CBVE}{\ensuremath{\pazocal{C}^{\mathrm{B}}_{\mathrm{v,e}}}\xspace}

\newcommand {\CNE}{\ensuremath{\pazocal{C}^{\mathrm{N}}_{\mathrm{e}}}\xspace}
\newcommand {\CSV}{\ensuremath{\pazocal{C}^{\mathrm{S}}_{\mathrm{v}}}\xspace}

\newcommand {\CCV}{\ensuremath{\pazocal{C}^{\mathrm{C}}_{\mathrm{v}}}\xspace}
\newcommand {\CCE}{\ensuremath{\pazocal{C}^{\mathrm{C}}_{\mathrm{e}}}\xspace}

\newcommand {\grd}{\ensuremath{\Delta^{\bullet}_{\mathrm{v/e}}}}

\newcommand\revise[1]{\textcolor{black}{#1}}

\makeatletter
\def\@email#1#2{%
 \endgroup
 \patchcmd{\titleblock@produce}
  {\frontmatter@RRAPformat}
  {\frontmatter@RRAPformat{\produce@RRAP{*#1\href{mailto:#2}{#2}}}\frontmatter@RRAPformat}
  {}{}
}%
\makeatother
\begin{document}


\title{A Perturbation-Based Approach to Identifying \revise{Potentially} Superfluous Network Constituents}
\author{Timo Br\"ohl}
\email{timo.broehl@uni-bonn.de}
\affiliation{Department of Epileptology, University of Bonn Medical Centre, Venusberg Campus 1, 53127 Bonn, Germany}
\affiliation{Helmholtz Institute for Radiation and Nuclear Physics, University of Bonn, Nussallee 14--16, 53115 Bonn, Germany}

\author{Klaus Lehnertz}
\affiliation{Department of Epileptology, University of Bonn Medical Centre, Venusberg Campus 1, 53127 Bonn, Germany}
\affiliation{Helmholtz Institute for Radiation and Nuclear Physics, University of Bonn, Nussallee 14--16, 53115 Bonn, Germany}
\affiliation{Interdisciplinary Center for Complex Systems, University of Bonn, Br{\"u}hler Stra\ss{}e 7, 53175 Bonn, Germany}

\date{\today}

\begin{abstract}
Constructing networks from empirical time series data is often faced with the as yet unsolved issue of how to avoid \revise{potentially} superfluous network constituents. Such constituents can result, e.g., from spatial and temporal oversampling of the system's dynamics, and neglecting them can lead to severe misinterpretations of network characteristics ranging from  global to local scale. We derive a perturbation-based method to identify \revise{potentially} superfluous network constituents that makes use of vertex and edge centrality concepts. We \revise{investigate the suitability of our approach through analyses of weighted small-world, scale-free, random, and complete networks}. 
\end{abstract}

\maketitle

\begin{quotation}
Understanding complex dynamical systems such as climate and brain profits from the network approach. Deriving networks from measurements of the systems' dynamics, however, can lead to spurious indications of network properties, depending on the employed sampling strategies and time-series analysis techniques to define networks constituents.
This, together with limitations in knowledge about the system's actual structural organization, calls for approaches to identify \revise{potentially} superfluous network constituents.
Here, we present such an approach.
It is based on minuscule and elementary perturbations targeting single network constituents.
Constituents are deemed \revise{potentially} superfluous if the perturbations lead to no or only negligible changes of network characteristics, covering the local to global scale.
We test our approach on various paradigmatic network models.
\end{quotation}

\frenchspacing

\section{\label{sec:Introduction}Introduction}
Complex network approaches have been repeatedly shown to provide deeper insights into structure and dynamics of spatially extended complex systems in diverse areas of science.
In many natural and man-made networked systems, access to the underlying coupling structure may be restricted or even impossible~\cite{boccaletti2006,arenas2008,bullmore2009,donges2009b,allen2011,barthelemy2011,barabasi2011,newman2012,baronchelli2013,lehnertz2014,heckmann2015,gao2016}.
Nevertheless, in such cases can the system be described by an interaction network with vertices representing subsystems or elementary units and edges representing interactions between them.
This ansatz has been successfully applied e.g. in the study of (functional) brain networks~\cite{bullmore2009,lehnertz2014}, climate networks~\cite{donges2009,Zhou2015}, protein-protein interactions~\cite{Uetz2000}, gene interactions~\cite{Tyler2009}, plant-pollinator interactions~\cite{Hegland2009,Olesen2011,halekotte2020}, food-webs~\cite{Delmas2019}, or communication and social networks~\cite{onnela2007,Palla2007}.

When it comes to analyzing real-world complex systems, lacking explicit knowledge of the structural organization of the dynamical system under study is quite common.
Hence, vertices of the related interaction network are commonly associated with sensors that are placed to sufficiently capture a subsystems' dynamics.
Deriving edges from the system's dynamics is usually based on a data-driven quantification of interaction properties, namely strength, direction, and coupling function.
Given that interactions can manifest themselves in various aspects of the dynamics (amplitudes, frequencies, phases, etc.), a large number of time series analysis techniques is now available.
The reliability of techniques, however, may be influenced by the mostly unavoidable finiteness of noisy field data which can lead to erroneous estimates of interaction properties.
Together with the fact that there is by now no commonly accepted method to derive binary or weighted (or weighted and directed) networks from interaction properties, this represents a source for severe misinterpretations of network properties~\cite{bialonski2010,hlinka2012,porz2014,wens2015,gastner2016,papo2016,hlinka2017,zanin2018}.

Yet, these issues are influenced and preceded by a more general problem: choosing the right number of sensors and placing them in a meaningful way.
Arrangement and placement of sensors is highly non-trivial and most often leads to a spatial over- or undersampling of a system.
These issues translate to the presence of additional and potentially superfluous constituents or the absence of potentially relevant constituents and may cause severe misinterpretations of network properties~\cite{bialonski2010,bialonski2011b,hlinka2012,chung2012,porz2014,wens2015,gastner2016,papo2016,hlinka2017,zanin2018}.
When investigating real-world systems, aiming to gather as much information as possible is rather common practice but bears the risk of oversampling the system.
Hence, there is a strong need for methods that allow to identify redundant or superfluous vertices and edges. 
For \revise{edges}, a vast plethora of methods has been proposed~\cite{lu2011,lue2015,liao2017,kramer2009,yan2018}, although their suitability continues to be matter of debate~\cite{zeng2012,zhang2018,kumar2020,cantwell2020}.
Interestingly, the issue of identifying superfluous vertices has so far been addressed only rarely~\cite{frantz2009,bellingeri2020b}.

We here propose a perturbation-based approach to identify \revise{potentially} superfluous network constituents (vertices and edges), employing elemental and minuscule perturbations that directly target single constituents.
With the premise that targeting constituents of \revise{potentially} superfluous nature has little to no effect on the \revise{characteristics of networks}, exactly these are compared prior and after perturbation.
\revise{We test the suitability of this method on weighted small-world, scale-free, random, and complete networks.}

\section{Methods}
There are several ways to perturb a network, with different types of perturbation potentially leading to different outcomes.
Almost all perturbations, however, can be viewed as an accumulation, superposition, or interplay of the following, straightforward and elementary perturbations:
\begin{enumerate}
    \item vertex/edge removal: one or more constituents are removed from the network. In case of removing a vertex, its connected edges are removed along with it. 
    The removal of constituents can have a significant impact on the network's \revise{connectedness}, and it can lead to the fragmentation of the network into smaller disconnected components;
    \item vertex/edge addition: constituents are added to the network. This can increase the network's \revise{connectedness} and can facilitate the exchange of information or resources in different ways;
    \item rewiring: an edge is randomly rewired, possibly leading to changes in the network's topology. Rewiring can alter the network's characteristics, especially path-structural aspects; 
    \item weight changes: the weight of an edge is altered, possibly influencing local up to global network characteristics. 

\end{enumerate}
However, the influences of structurally minuscule perturbations are hard to determine and even harder to control on the level of the complex system.
Previous research has shown that random perturbations can have major influence on very specific macroscopic network \revise{characteristics}~\cite{Latora2005,Ghoshal2011,Ceci2018}.
Nonetheless, and especially when it comes to the investigation of real-world systems, it remains unclear, how on a general basis minuscule perturbations targeting single constituents change the respective networks.
Thus far, comparing networks is a notorious difficult task, particularly for networks of different sizes (and changes in network sizes often go hand in hand with these perturbations) and there is no commonly accepted and sufficient way to do so~\cite{Tantardini2019,Mheich2020}.
Hence, we can only focus on network \revise{metrics}~\cite{rings2022} that, in total, describe the network somewhat comprehensively (cf. Sect.~\ref{sec:network_metrics} and Table~\ref{tab:measures_metrics}).

The principal idea behind our perturbation-based approach now is, that if the targeted perturbation \revise{of a network constituent} (cf. Sect.~\ref{sec:perturbation_method}) does not alter \revise{network characteristics assessed with the various metrics, or only to a small negligible extent, the targeted constituent can be deemed potentially superfluous}.

\subsection{Employed perturbations}
\label{sec:perturbations}
Of the above listed four elementary perturbations, only the first two are universally applicable in any kind of network, independent of its topology and definitions of edges (regarding weight, direction or multiple edges).
Based on these, we further differentiate between the following three perturbations that we will employ in the wake of our perturbation-based method to identify \revise{potentially} superfluous constituents:
\revise{
\begin{itemize}
    \item vertex removal: a vertex $v$ and its connected edges $\{v, j\}$ are discarded, with $j$ denoting vertices adjacent to $v$;
    \item vertex cloning: a vertex $v$ --~that is already present in the network~-- is duplicated, including its connected edges $\{v, j\}$ 
    by adding a vertex $v'$ and adding the respective edges $\{v', j\}$, with $j$ denoting vertices adjacent to $v$; cloned vertex $v$ and clone $v'$ are not directly connected;
    \item edge removal: an edge is discarded from the network.
\end{itemize}
}
Perturbations directly targeting a single vertex, hence may indirectly affect edges connected to the respective vertex in the course of the perturbation.

\subsection{\revise{Network metrics}}
\label{sec:network_metrics}
\revise{We generally differentiate between global and local network characteristics that are evaluated by network metrics (cf. Table~\ref{tab:measures_metrics})}. 
Global network \revise{metrics describe the network as a whole, often associating a network characteristic with a single quantity.}
Local network \revise{metrics} focus on aspects of single network constituents (vertices and edges) or groups of such.
While some of these local \revise{metrics} might still depend on the composition of the network as a whole, others merely depend on the direct neighborhood of the respective constituent. 

A network consists of \revise{a set of vertices $\mathcal{V}$ ($v_i\in\mathcal{V}, i=1,\dots,V$; $V=|\mathcal{V}|$) and a set of edges ($e_n\in\mathcal{E}, n=1,\dots,E$; $E=|\mathcal{E}|$) with an edge connecting two vertices ($e_{ij}={v_i,v_j}$).}
The network can be described by its adjacency matrix $\mathcal{A}\in\{0,1\}^{V\times V}$, with $\mathcal{A}_{ij}=1$ if edge $e_{ij}$ exists between vertices $i$ and $j$, and $\mathcal{A}_{ij}=0$ otherwise.
\revise{Complementarily, for weighted networks}, we define the weight matrix \revise{$\mathcal{W}\in[0,1]^{V\times V}$}, with $\mathcal{W}_{ij}$ referring to the edge weight (strength of interaction) between vertices $i$ and $j$.

\subsubsection{Global network \revise{metrics}}
\revise{Some metrics of networks describe certain aspects of the network as a whole.}
This not only can allow the classification of network topologies, but also allows the comparison of other characteristics of networks, such as network size, path-structure, degree-correlations, robustness, and stability.

The {\bf (pseudo) diameter} $D$ is the length of the longest shortest path between any pair of vertices in a network.
\revise{The length of a path is chosen as the sum of the inverse of all edge weights on that path.}\\

\revise{
The {\bf average shortest path length} $L$ quantifies the average length of a path $\psi$ between any two vertices ($\{z,l\}\in\mathcal{V}$) in a network:
\begin{equation*}
	L=\frac{\sum_{zl}\psi_{zl}}{V(V-1)}.
\end{equation*}
}

The {\bf global clustering coefficient} $G$ quantifies to what extend network vertices tend to cluster together. 
For a weighted network, the global clustering coefficient is defined as:
\begin{equation*}
G=\frac{\mathrm{Tr}\mathcal{W^3}}{\sum_{z\neq l}[\mathcal{W}^2]_{zl}}.
\end{equation*}

{\bf Assortativity} $A$ characterizes how vertices with (dis\nobreakdash)similar properties (here: strength, \revise{being the sum of the weights of the attached edges}) are preferentially connected amongst themselves~\cite{Newman2002, Bialonski2013}.
To calculate $A$, we estimate the (Pearson) correlation coefficient between the strengths of connected vertices:
\begin{equation*}
	A=\sum_{xy}xy(q_{xy}-a_xb_y)/\sigma_a\sigma_b,
\end{equation*} 
with $x$ and $y$ representing strength values and $q_{xy}$ representing the fraction of edges that connect a vertex with strength $x$ to a vertex of strength $y$.
Then $q_{xy}$ satisfies the sum rules: $\sum_{xy}q_{xy}=1$, $\sum_{y}q_{xy}=a_x$, $\sum_{x}q_{xy}=b_y$.
$\sigma_a$ and $\sigma_b$ are the standard deviations of the distributions $a_x$ and $b_y$.\\

{\bf Synchronizability} $S$ of a network describes the stability of its globally synchronized state~\cite{Barahona2002, Atay2006}.
We here characterize it by the eigenratio $S=\lambda_V/\lambda_2$. 
$\lambda_V$  denotes the largest eigenvalue and $\lambda_2$ the smallest non-vanishing eigenvalue of the network's Laplacian matrix $L_{zl}=s_z\delta_{zl}-\mathcal{W}_{zl}$ ($\delta$ is the Kronecker delta, $s_z$ denotes the strength of vertex $z$; see below).

\subsubsection{Local network \revise{metrics}}
The concept of centrality has been introduced in many different fields and contexts~\cite{beauchamp1965,Sabidussi1966,freeman1977,freeman1979,bonacich1987,Wuchty2003,koschutzki2005,borgatti2006,estrada2010,Valente2010,Chen2012,Costa2015,lawyer2015,Wu2018,broehl2019,Zhao2020,Broehl2022}. 
The general idea is to quantify a constituent's role or importance in the larger network based on certain characteristics, primarily focusing on the integration of a constituent in the network due to specific aspects. 
Hence centrality \revise{metrics can be utilized to measure} importance yielding an importance ranking~\cite{Ghoshal2011,lue2016,Iniguez2022}.\\ 

The {\bf degree} of a vertex $z$ is the sum of edges connected to this vertex: $d_z=\sum_{l=0}^{V} \mathcal{A}_{zl}$.
Analogously the {\bf strength} \revise{(or {\bf strength centrality})} of a vertex is the sum of edge weights of all edges connected to this vertex: \revise{$s_z=\CSV(z)=\sum_{l=0}^{V} \mathcal{W}_{zl}$.}\\

Similarly, with {\bf nearest-neighbor centrality}~\cite{Broehl2022}, an edge is considered to be more central the larger its weight and the more similar and the higher the strengths of the vertices which are connected by that edge.
Nearest-neighbor edge centrality of an edge $z$ between vertices $a$ and $b$ is defined as~\cite{Broehl2022}
\begin{equation*}
\CNE(z)=\frac{\CSV(a)+\CSV(b)-2w_z}{|\CSV(a)-\CSV(b)|+1}\;w_z,
\label{eq:CNE}
\end{equation*}
where $w_z=\mathcal{W}_{ab}$ denotes the edge weight and $z \in \left\{1,\ldots,E\right\}$ and $(a,b)\in\{1,\dots,V\}$. 
Hence, nearest-neighbor centrality can be considered as a strength-based edge centrality concept.
Much like the strength of a vertex, the nearest-neighbor centrality value of an edge is only influenced by its adjacent constituents. \\

{\bf Eigenvector centrality} considers the influence of a vertex/edge (v/e) on the network as a whole. 
A constituent is regarded as central if adjacent constituents are also central. This centrality is defined as~\cite{bonacich1972,broehl2019}
\begin{equation*}
\CEVE(z)=\frac{1}{\lambda_{\max}}\sum_{l}^{}M_{zl}\,\CEVE(l).
\label{eq:CEVE}
\end{equation*} 
In case of vertices, $\left\{z,l\right\}\in\mathcal{V}$ and ${\bf M}$ denotes the weight matrix $\mathcal{W}^{\rm (v)}\in[0,1]^{V\times V}$, with $\mathcal{W}^{\rm (v)}_{zl}$ denoting the weight of an edge between vertices $z$ and $l$. 
We define $\mathcal{W}^{\rm (v)}_{zz} \defi 0 \,\forall\, z$ with $z\in\left\{1,\ldots,V\right\}$.
In case of edges, $\left\{z,l\right\}\in\mathcal{E}$ and ${\bf M}$ denotes the weight matrix $\mathcal{W}^{\rm (e)}\in[0,1]^{E \times E}$ whose entries $\mathcal{W}^{\rm (e)}_{zl}$ are assigned the average weight of edges $z$ and $l$ if these edges are connected to a same vertex, and 0 otherwise. 
As above, we define $\mathcal{W}^{\rm (e)}_{zz} \defi 0 \,\forall\, z$ with $z\in\left\{1,\ldots,E\right\}$.
The aforementioned definition is applied iteratively until eigenvector centrality values remain stable.
Eigenvector centrality can be considered as a strength-based centrality concept, which, due to its recursive definition, relates a constituent to all the other constituents in the network.\\

{\bf Closeness centrality} considers the distance between a constituent and all other constituent in a network. 
A constituent with a high closeness centrality is central as information from this constituent can reach all other constituents in the network via short paths, and so the constituent can exert a more direct influence over the network. 
Closeness centrality of vertex $z$ is defined as\cite{bavelas1950}:
\begin{equation*}
\CCV(z)=\frac{V-1}{\sum_{l}{d_{zl}}}, 
\label{eq:CCV}
\end{equation*}
with $(z,l) \in \left\{1,\ldots,V\right\}$ and where $d_{zl}$ is the length of the shortest path between vertices $z$ and $l$, calculated as the sum of the inverse of all edge weights on the path.
Closeness centrality of edge $z$ between vertices $a$ and $b$ can be defined as\cite{broehl2019}:
\begin{equation*}
\begin{aligned}
\CCE(z) & =\frac{E-1}{\sum_{l}{(d_{la}+d_{lb})}} 
=\frac{E-1}{\frac{1}{\CCV(a)}+\frac{1}{\CCV(b)}}\\
& =(E-1)\frac{\CCV(a)\CCV(b)}{\CCV(a)+\CCV(b)}, 
\label{eq:CCE}
\end{aligned}
\end{equation*}
with $z \in \left\{1,\ldots,E\right\}$ and $(a,b,l)\in\{1,\dots,V\}$.
Hence, closeness centrality can be considered as a path-based centrality concept, which is therefore influenced by the network as whole.\\

{\bf Betweenness centrality} is a measure of how frequently a shortest path traverses a given constituent.
A constituent with a high betweenness centrality value is central because it acts as a bridge between other parts of the network. 
Vertex/edge betweenness centrality (of vertex/edge $z$) can be defined as~\cite{freeman1977,brandes2001,girvan2002,broehl2019}
\begin{equation*}
\CBVE(z)=\frac{2}{F}\sum_{l\neq m}\frac{q_{lm}(z)}{G_{lm}}, 
\label{eq:CBVE}
\end{equation*}
where $z \in \left\{1,\ldots,V\right\}$ (for vertices), resp. $z \in \left\{1,\ldots,E\right\}$ (for edges), $\left\{l,m\right\} \in \left\{1,\ldots,V\right\}$, $q_{lm}(z)$ is the number of shortest paths between vertices $l$ and $m$ running through vertex/edge $z$, and $G_{lm}$ is the total number of shortest paths between vertices $l$ and $m$.
\revise{Again, the length of a path is chosen as the sum of the inverse of all edge weights on that path.} 
The normalization factor is $F=(V-1)(V-2)$ in case of vertices and $F=V(V-1)$ in case of edges.
Betweenness centrality can be considered as a path-based centrality concept, which is therefore influenced by the network as whole.\\

In order to be able to compare results yielded by different centrality concepts qualitatively, we introduce a centrality-value-based ranking of the networks constituents. 
A vertex/edge is assigned rank 1 if the largest centrality value is associated with it.
The rank increases in increments of 1 for the vertex/edge with second largest centrality value, third largest centrality value etc., yielding an increasing rank as centrality values decrease.
We abstain from assigning
two or more constituents the same rank and rank in order of
appearance for equal centrality values.

\begin{table*}[htbp]
	\resizebox{0.8\textwidth}{!}{
	\def\arraystretch{1.2}\tabcolsep=5pt
	\revise{
	\begin{tabular}{ll|p{7cm}|l}
		&                           & characteristics                                                                                        & metrics                           \\ \hline
		\multirow{9}{*}{global} & \multirow{9}{*}{network}  & length of the longest shortest path between any pair of vertices in a network                          & (pseudo) diameter $D$                \\
		&                           & average length of a path between any two vertices in a network                                         & average shortest path length $L$  \\
		&                           & extent to which vertices in the network tend to cluster together                                       & global clustering coefficient $G$ \\
		&                           & extent to which vertices with (dis-)similar properties are preferentially connected amongst themselves & assortativity $A$                 \\
		&                           & stability of the network’s globally synchronized state                                                 & synchronizability $S$           \\ \hline
		\multirow{17}{*}{local}  & \multirow{10}{*}{vertices} & intergration of a vertex in its direct neighborhood (binary network)                                   & degree $d$                        \\
		&                           & intergration of a vertex in its direct neighborhood (weighted network)                                 & strength centrality $\CSV$        \\
		&                           & extent to which a strongly integrated vertex is connected to other strongly integrated vertices        & eigenvector centrality $\CEV$     \\
		&                           & extent to which a vertex can reach any other vertex in the network via shortest paths                  & closeness centrality $\CCV$       \\
		&                           & extent to which a vertex connects otherwise distant regions in the network                             & betweenness centrality $\CBV$     \\ \cline{2-4} 
		& \multirow{7}{*}{edges}    & integration of an edge in its direct neighborhood                                                      & nearest-neighbor centrality $\CNE$           \\
		&                           & extent to which a strongly integrated edge is connected to other strongly integrated edges             & eigenvector centrality $\CEE$     \\
		&                           & extent to which an edge can reach any other edge in the network via shortest paths                     & closeness centrality $\CCE$       \\
		&                           & extent to which an edge connects otherwise distant regions in the network                              & betweenness centrality $\CBE$     \\ \hline
	\end{tabular}
	}
	}
\revise{
\caption{Overview of global and local network characteristics and their respective quantifying metrics. \label{tab:measures_metrics}}
}
\end{table*}

\revise{\subsection{Perturbations of network constituents}
\label{sec:perturbation_method}
For a given realization of a network (out of $N_{\rm r}$ realizations),
\begin{itemize}
	\item we estimate local and global network metrics (cf. Sect.~\ref{sec:network_metrics})
	and deduce an initial ranking of the network's constituents based on their centrality values.
	The latter allows to identify a constituent, based on its rank, in the different realizations;
	\item we iterate over all network constituents and
	\begin{itemize}
		\item employ the respective perturbation (cf. Sect.~\ref{sec:perturbations}) to the test if the respective constituent $\gamma$ is potentially superfluous, 
		\item estimate local and global network metrics for this perturbed network,
		\item quantify the influence of the perturbation by comparing local and global network metrics prior and after the perturbation (cf. Sect.~\ref{sec:quantifying}).
	\end{itemize}
\end{itemize}
}

\subsection{Quantifying influences of network perturbations}
\label{sec:quantifying}
In case of the global network \revise{characteristics}, we \revise{track the perturbed} constituent $\gamma$ throughout the realizations of a given network via its respective rank $r_{\rm u}(\gamma)$ \revise{prior to perturbation (estimated with \CSV for vertices and with \CNE for edges)}. 
Generally, we expect the structurally minuscule perturbations to also have a negligible impact on any global network \revise{characteristics}, if such does not strongly depend on the number of vertices or the number of edges.
We calculate the average percentage change (from $N_{\rm r}$ realizations) of each \revise{metric} resulting from the perturbation as\revise{
\begin{equation*}
	\overline{\delta \mu^{(r_{\rm u}(\gamma))}} =\frac{1}{N_{\rm r}}\sum_{i=1}^{N_{\rm r}}100\frac{ \mu_{i, \rm u}-\mu_{i, \rm p}^{(r_{\rm u}(\gamma))}}
	{\mu_{i, \rm u}}.
\end{equation*}
Here $\mu\in\{D,L,G,A,S\}$, $\mu_{i,\rm u/p}$ denotes the global metric of the unperturbed/perturbed network (of realization $i$) and $r_{\rm u}(\gamma) \in \left\{1,\ldots,V\right\}$ (for vertices), resp. $r_{\rm u}(\gamma) \in \left\{1,\ldots,E\right\}$ (for edges) is the rank of the perturbed constituent $(\gamma)$ in the unperturbed network.}

In case of the local network characteristics and when investigating a change in centrality values, it is important to recognize that there is no one true centrality concept.
Each of the centrality concepts employed here does focus on different topological aspects of the network. 
However, we can generally divide the centrality concepts into strength-based (nearest-neighbor centrality and eigenvector centrality) and path-based (betweenness centrality and closeness centrality) concepts.
\revise{Hence, it is not necessarily to be expected to observe perturbation-induced changes in the distribution of strength-based centrality values, when observing a perturbation-induced change in the distribution of path-based centrality values (or vice versa)}. 
Nevertheless, in order for a targeted constituent to be \revise{considered potentially} superfluous, the \revise{respective} perturbation should not lead to meaningful changes in either \revise{distribution}. 
\revise{For the employed perturbations, we would expect largely comparable distributions of centrality values for a given network prior and after perturbation (independent of the centrality concept)}.
\revise{The values of a given centrality metric, for a given network prior to and after perturbation, can therefore be considered to be drawn from the same distribution, and we test this null hypothesis using the Kolmogorov-Smirnov test.} 
The null hypothesis is rejected for $p<0.05$.
Hence, if the null hypothesis can be rejected for a certain perturbation, and with regard to any of the centrality concepts, the perturbed network constituent can not be considered \revise{potentially} superfluous under this perturbation.
However opposite reasoning, deeming a \revise{constituent as potentially superfluous is not valid if the respective perturbation did not lead to significant changes between the distributions of centrality values of the unperturbed and perturbed network.}
Still it can be a further indication of the targeted constituent being \revise{potentially} superfluous.
\revise{Nonetheless,} the specific local changes in the importance ranking of constituents can be abundant and meaningful in a greater context, while the distribution of centrality values is left unaltered. 

\revise{In case of the importance hierarchies, as deduced from the centrality-based rankings of a networks' constituents (vertices and edges, respectively)}, we quantify the local impact of a perturbation of constituent $\gamma$ by considering the following metric, calculating the difference \grd between the ranks~\cite{Saavedra2011} $r_{\rm u}(z)$ and $r_{\rm p}(z)$ of any constituent \revise{$z\neq\gamma$} ($z \in \mathcal{V}$ for vertices (v) and $z \in \mathcal{E}$ for edges (e)) for the unperturbed and the perturbed network:
\begin{equation*}
\grd (z)= |r_{\rm u}(z)-r_{\rm p}(z)|, 
\end{equation*}
where $\bullet \in\{\text{S,E,C,B}\}$ denotes the centrality concept employed for the ranking.
\revise{$\grd (z) \to 0$ can be considered as a further indication for constituent $\gamma$ to be potentially superfluous.}
The median value of $\grd (z)$ is expected to be rather small.

Overall, deeming a network constituent \revise{potentially} superfluous can not be considered an absolute truth, but is rather an assessment of a number of indications based on observed changes \revise{in network characteristics due to specific perturbations}.
The more qualifying observations can be made \revise{regarding} these network characteristics --~for the respective perturbation~-- the more considerable is a potential superfluous nature of the targeted constituent.

We thus set three criteria that indicate a constituent targeted by the respective perturbations to be considered \revise{potentially} superfluous: negligible changes in a number of \revise{global network metrics}, negligible changes in the distributions of centrality values (based on multiple centrality concepts), and negligible changes in the actual \revise{importance hierarchies} of the constituents.

\revise{With regard to potentially dependencies of the aforementioned criteria such as network topology, size and edge density, and thus with an eye on potential applications in the analyses of real-world systems, we investigate the suitability of these criteria analyzing various networks with preset properties.}

\section{Networks investigated}
When approximating real-world complex systems with networks, the latter are often associated with specific topologies (or combinations of such).
Independent of the underlying complex systems, these topologies can have quite distinct properties that may not only influence network characteristics substantially but also can induce 
superfluous constituents, at least in some of their realizations.
\revise{As an example, it may naively to be expected to find superfluous information in a very dense and large network (e.g., large random networks) while it is rather less likely to find such in sparse networks (e.g., small-world or scale-free networks) or in networks with regular structures (e.g., lattices or rings).}

We here consider undirected, weighted and connected networks without self-loops or multiple edges.
In the following, edge weights are drawn from a uniform distribution $U(0,1)$.
We investigate four different topologies and networks of different sizes ($V\in\{20,50,100,200,500,1000\}$) and different edge densities.
\revise{
Each network type of the following four topologies will be realized $N_{\rm r}=100$ times:
\begin{itemize}
	\item small-world networks~\cite{watts1998} with rewiring probabilities $p\in\{0.01,0.1,0.2,0.3\}$ starting from a ring with $m\in\{4,8\}$ nearest-neighbors being connected,
	\item random networks~\cite{erdos1959,batagelj2005} with edge creation probabilities $q\in\{0.05,0.1,0.2,0.3,0.5,0.7\}$,
	\item scale-free networks~\cite{albert2002} with the newly added vertices preferentially attached to existing vertices of high degree with $k\in\{4,6,10\}$ edges, 
	\item complete networks.
\end{itemize} 
The edge density for a network is then defined as:
\begin{equation*}
	\varrho=E/E_{\mathrm{max}}=E/\binom{V}{2}=\frac{2E}{V(V-1)}.
\end{equation*}
}

\section{Results}
\subsection{Impact of perturbations on \revise{global network characteristics}}
\label{criteria1}
For each of \revise {the three types of perturbation and predominantly independent of the network topology, we observe --~overall and on overage~-- perturbation-induced changes of global network metrics in the order of a few percent ($<5\,\%$, cf. Table \ref{tab:global})}.
\begin{table*}[htbp]
\resizebox{\textwidth}{!}{
\begin{tabular}{@{}l|lll|lll|lll|lll|lll@{}}
\toprule
\backslashbox[40mm]{perturbation}{\revise{metrics}}
               & $<\overline{\delta D}>$ & $\sigma(\overline{\delta D})$ & $\max(|\delta D|)$ & $<\overline{\delta L}>$ & $\sigma(\overline{\delta L})$ & $\max(|\delta L|)$ & $<\overline{\delta G}>$ & $\sigma(\overline{\delta G})$ & $\max(|\delta G|)$ & $<\overline{\delta A}>$ & $\sigma(\overline{\delta A})$ & $\max(|\delta A|)$ & $<\overline{\delta S}>$ & $\sigma(\overline{\delta S})$ & $\max(|\delta S|)$ \\
               \midrule
vertex removal & -1.5           & 4.85        & 16.66     & -0.81          & 1.47        & 29.33     & -0.19          & 2.01        & 15.93     & -1.01          & 12.44       & 39.63     & -4.87          & 19.83       & 1047.42   \\
vertex cloning & -0.01          & 0.06        & 1.76      & 0.03           & 0.24        & 5.69      & 0.91           & 1.55        & 9.73      & -0.37          & 13.74       & 53.97     & -1.62          & 3.23        & 21.8      \\
edge removal   & -0.56          & 2.02        & 9.46      & -0.42          & 0.71        & 28.37     & 0.37           & 1.19        & 6.65      & -0.39          & 4.6         & 24.98     & -1.72          & 8.73        & 1047.74   \\ \bottomrule
\end{tabular}
}
\caption{\label{tab:global} Percentage increase/decrease (as well as standard deviation and maximum value) of \revise{global network characteristics under respective perturbation and averaged over all constituents and all investigated networks ($<\bullet>$).}
$D$ diameter; $L$ average shortest path length; $G$ global clustering coefficient; $A$ assortativity; $S$ synchronizability.}
\end{table*}
It is to be mentioned that while, on average, pointing to rather negligible changes, we do observe large maximum changes in few, specific \revise{networks} --~primarily observed \revise{for} assortativity (up to $\sim 50\,\%$) and synchronizability (up to $\sim 1000\,\%$)~-- \revise{the latter} pointing towards the existence of some rare but seemingly vital constituents regarding the stability of a synchronized state particularly in random and complete networks. 

\begin{figure}[htbp]
\centering
\includegraphics[width=1\linewidth]{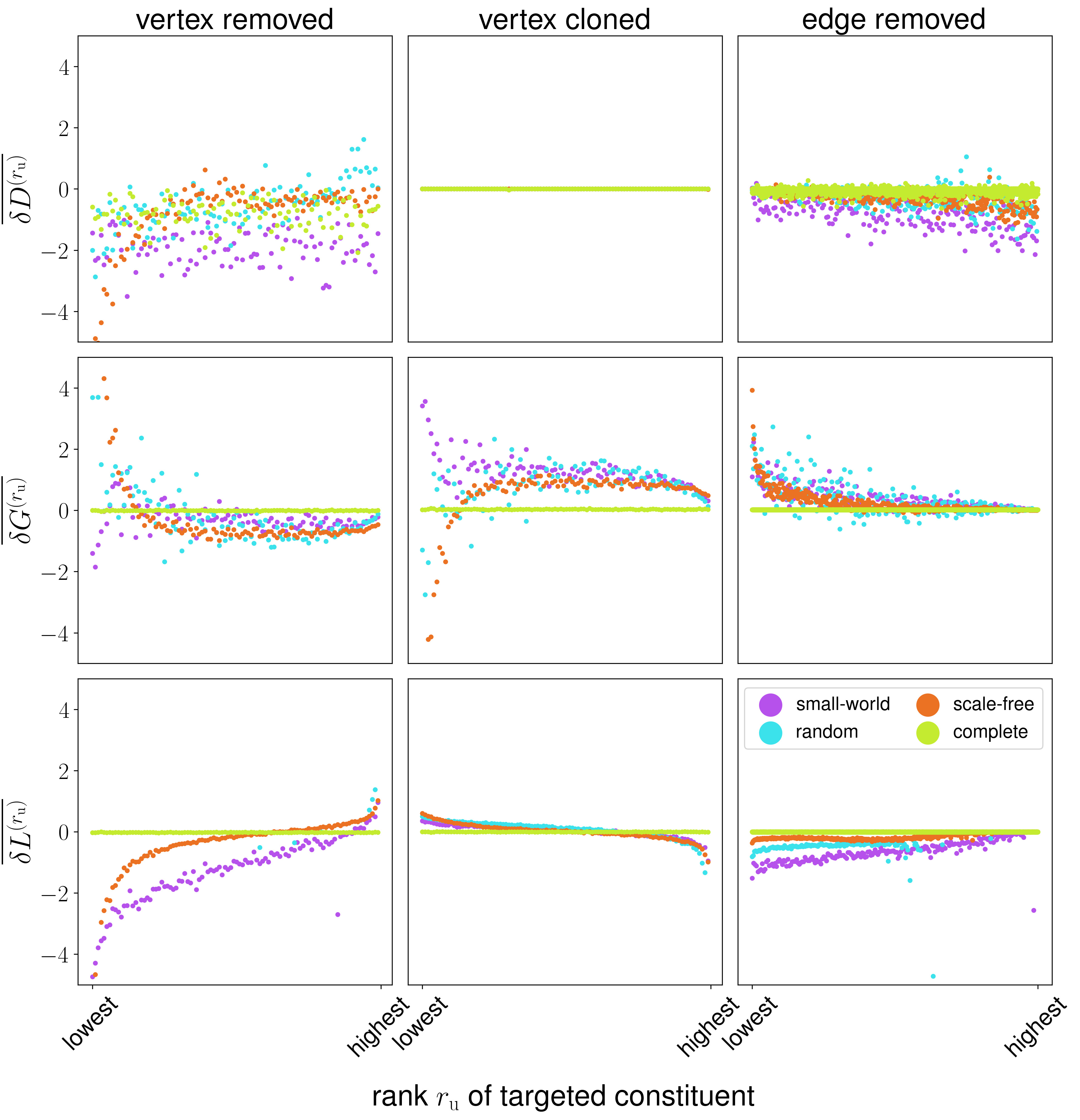}
\caption{Averages of percentage changes of (pseudo) diameter $\overline{\delta D^{(r_{\rm u})}}$, of global clustering coefficient  $\overline{\delta G^{(r_{\rm u})}}$, and of average shortest path length  $\overline{\delta L^{(r_{\rm u})}}$ under the respective perturbation \revise{of a targeted constituent with certain rank $r_{\rm u}$}. Vertex ranks estimated with strength centrality and edge ranks estimated with nearest-neighbor centrality \revise{prior to perturbation}. Averages from 100 realizations each of four exemplary network types (small-world network: $V=100$, $m=4$, $p=0.2$ --- random network: $V=100$, $q=0.05$ --- scale-free network: $V=100$, $k=4$ --- complete network: $V=100$).}
\label{fig:global_exmaples}
\end{figure}

\revise{Furthermore, we do observe dependencies regarding the importance $r_{\rm u}$ of the constituent targeted by the perturbation, for at least some of the global network metrics}.
Yet, these dependencies vary regarding the network topology.
Especially for those network topologies that are less random and of more regular structure (small-world and scale-free networks) the magnitude of changes of global clustering coefficient and average shortest path length (but also of assortativity and synchronizability; data not shown) depend on the importance \revise{$r_{\rm u}$} of the removed/cloned constituent.
Generally, largest total changes \revise{of metrics} can be observed when removing/cloning most and least important constituents (cf. Figure \ref{fig:global_exmaples}).

\revise{
Surprisingly, we also observe not the removal of an edge but the cloning of a vertex to have smallest average impact on four of the five \revise{global network metrics}, even though removing a single edge is arguably a smaller structural network perturbation. 
It can be concluded that the employed network perturbations overall lead to minuscule changes of global network characteristics.
Nevertheless, it is vital to realize that the observe changes can depend on the targeted constituents' importance.
\revise{Our results indicate that particularly those constituents at neither end of the importance ranking can be deemed potentially superfluous.}}

\subsection{Impact of perturbations on \revise{local network characteristics}}
\label{criteria2}
The observed dependencies of changes of global network characteristics regarding the \revise{targeted constituent's importance $r_{\rm u}$} and the type of perturbation indicate that similar changes and dependencies can be observed on smaller network scales as well.
In case of local network \revise{metrics}, we first investigate whether deviations in the distribution of centrality values --~for any of the four centrality concepts (vertex and edge centralities, respectively)~-- can be observed after perturbation (cf. Sect.~\ref{sec:quantifying}).
If centrality values from an unperturbed and perturbed network can generally be considered to be drawn from different distributions, the removed/cloned vertex/edge can hence not be deemed \revise{potentially} superfluous.
Naively viewed, it is still apparent that of the three perturbations, removing a single edge is the smallest structural network alteration, as removing/cloning a vertex would also include the removal/cloning of attached edges.
Hence, it is generally to be expected that removing an edge has not only the smallest impact but potentially no impact at all on the distribution of centrality values compared to removing or cloning a vertex.
\begin{figure*}[htbp]
\centering
\includegraphics[width=\linewidth]{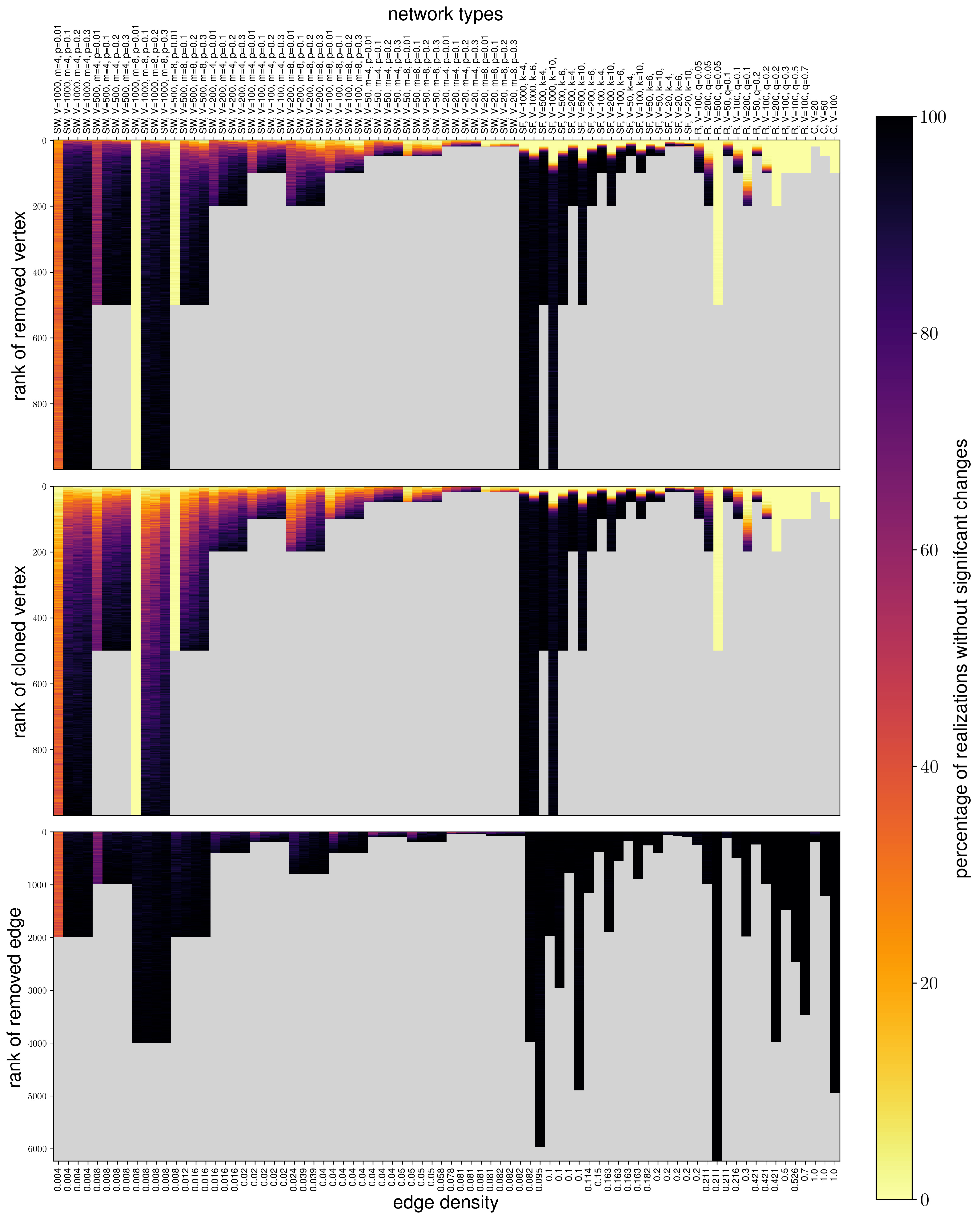}
\caption{Percentage of network realizations (color-coded) that show no significant change in either of the centrality distributions (cf. Section \ref{sec:quantifying}) when removing/cloning a constituent of a given rank $r_{\rm u}$ (estimated with strength/nearest-neighbor centrality, in the respective network) (SW small-world network; SF scale-free network; R random network; C complete network).
Empty cells -- due to differences in network size -- are colored in grey.}
\label{fig:cent_distribution_signficant}
\end{figure*}

Figure~\ref{fig:cent_distribution_signficant} shows that the respective perturbations did not lead to a significant change in the distribution of centrality values in the majority of investigated networks. 
\revise{We do, however, observe the changes to strongly depend on network topology, type of perturbation, and rank $r_{\rm u}$ of removed/cloned constituent.}
In case of removing an edge, we only observe very few significant changes whatsoever in small-world networks with small rewiring probability ($p=0.01$).
\revise{Their almost regular structure explains why removing only important edges still can lead to changes in the distribution of centrality values in at least some of the realizations of networks (<$40\%$), as removing such an important edge from the network will lead to large changes in the path-structure and thus will greatly affect centrality values estimated with path-based centrality metrics.}
Similar effects, that can be explained in an analogous way, regarding these specific small-world networks (the 12 small-world network types with $p=0.01$) are observed when removing or cloning a vertex.
\revise{The more regular the network the stronger is the alteration of the regular structure when introducing or removing a vertex and its respective edges.
Furthermore, as the ranking in case of the vertices is done via the strength centrality, it can be deduced that the more important the vertex, the larger is its integration in its direct neighborhood within the network (cf. Table \ref{tab:measures_metrics}).
Therefore the more important the removed/cloned vertex, the larger the alteration of the network's structure.}

Especially \revise{the aforementioned small-world networks} that are large in size and have a high edge density ($V \in\left\{500, 1000\right\}$, $m=8$) show the largest amount of significant changes.
This is, otherwise, only observed for networks with largest possible edge densities (fully connected networks) or comparably large random networks ($V=500$).
For these large and/or dense networks, cloning a vertex will consequently result in adding a large amount of edges\revise{, due to cloned vertex' high degree.}
Likewise, removing a vertex includes removing a large amount of edges in these networks. 
Both perturbations hence result in large changes of the distributions of centrality values.

\revise{In case of the less regular and less dense networks (random networks with $V<500$, small-world networks with $0.01<p<1.0$, scale-free networks), the observed changes in centrality values highly depend on which vertex was removed/cloned.}
The amount of network realizations \revise{with} significant changes in the distributions of centrality values (regarding the respective perturbation) decreases with increasing rank (decreasing importance) of the targeted vertex.
\revise{This once again is likely explained by the high interconnectedness (high degree) of important vertices.}
\revise{Cloning the most important vertex always -- in $100\,\%$ of the realizations -- led to significant changes of the distribution of centrality values with regard to the unperturbed network.
Furthermore, and especially in small-world networks, even cloning less important vertices much more often led to significant changes than removing said vertices.}

We can conclude, that the structurally smallest perturbation, namely removing a single edge, has the overall smallest influence on the distribution of centrality values and that removing a vertex less often leads to significant changes than cloning said vertex.
Furthermore, removing/cloning an important constituent has a stronger impact than removing a less important constituent.
Almost regular as well as dense and large networks are most strongly affected by perturbations targeting vertices.
\revise{Our results here, and in contrast to those reported on in Sect.~\ref{criteria1}, indicate that particularly those constituents identified as less important prior to perturbation (high rank), may qualifies as potentially superfluous.} \\

\subsection{Impact of perturbations on importance hierarchy}
\label{criteria3}
\revise{Having observed mostly insignificant changes in the distribution of centrality values, at first glance, points toward a weak and minimal alteration of the network.
Yet, the importance hierarchies of network constituents might have changed greatly.
As a most extreme example: the constituents with respectively highest and smallest centrality value prior to perturbation exchange their positions in the ranking as a result of the perturbation.
A constituent with little importance prior to perturbation is now, after the employed perturbation, considered the most important constituent and vice versa, while the distribution of centrality values remained the same.
Furthermore, changes in the importance hierarchies can be used to identify if constituents that are directly affected by the perturbation (e.g., removed or cloned) can be deemed \revise{potentially} superfluous.  
To this end, we investigate the changes in constituents ranks due to the respective perturbations (cf. Sect.~\ref{sec:quantifying}).}

\begin{figure*}[htbp]
\centering
\includegraphics[width=\linewidth]{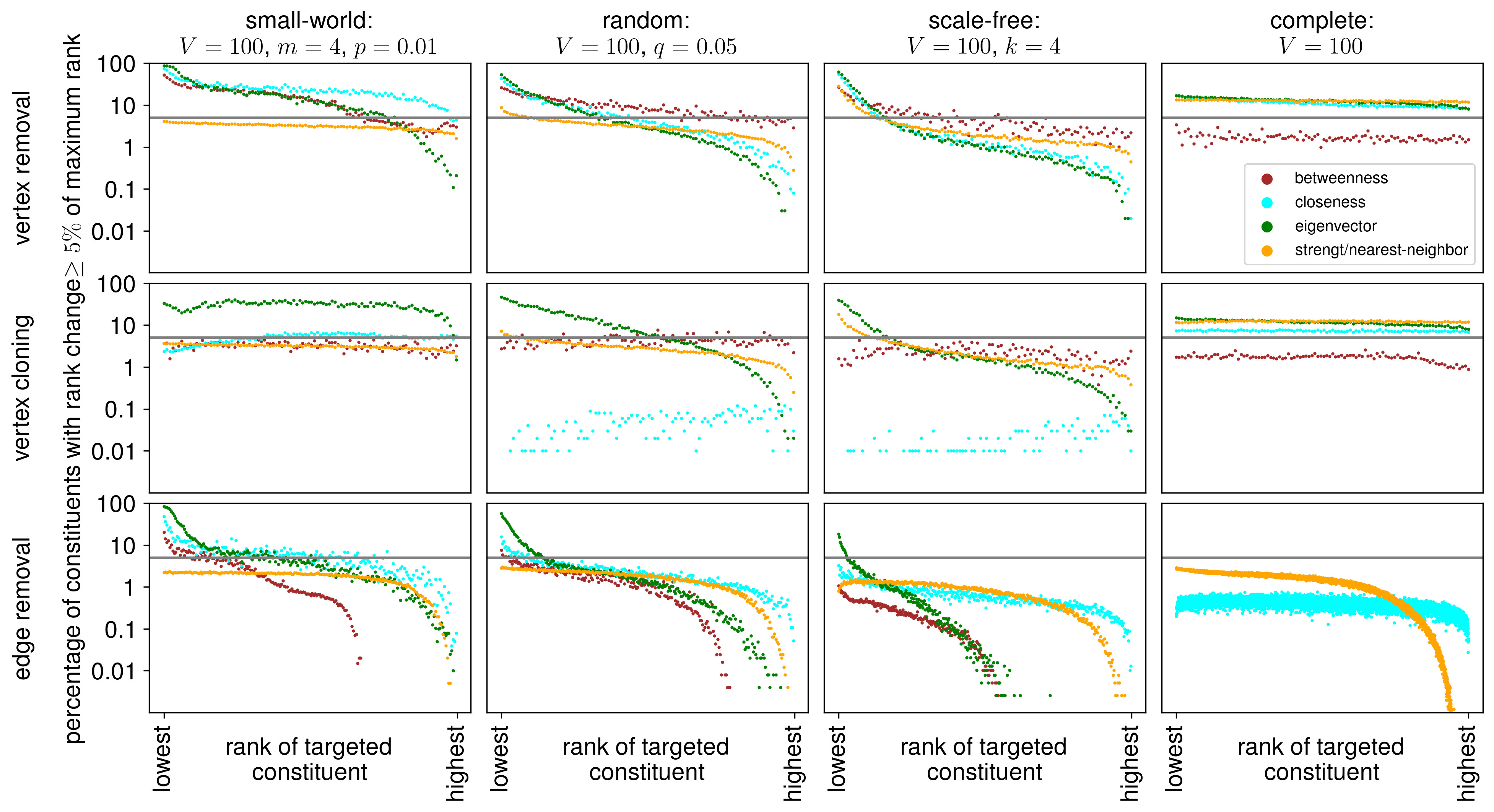}
\caption{\revise{Percentage of constituents with rank changes $\geq\theta$ of the maximum possible rank change in dependency of the rank of the constituent targeted by the respective perturbation (rows).
We here chose $\theta\geq5\%$ as this amounts to $\grd\geq1$ in case of the smallest investigated networks with $V=20$.}
Ranking is done via different centrality concepts (color-coded). Data is averaged over 100 realizations  of each of four exemplary network types (columns): small-world network ($V=100$, $m=4$, $p=0.2$), random network ($V=100$, $q=0.05$), scale-free network ($V=100$, $k=4$), and complete network ($V=100$). The black horizontal line is for eye guidance only and represents $5\%$ of \revise{possible constituents showing a rank change $\grd\geq1$}. Similar relations are observed for $20\geq V \geq 1000$.}
\label{fig:fig3}
\end{figure*}
It can be deduced from Figure \ref{fig:fig3} that, largely independent of the employed perturbation, the changes in the importance hierarchies depend on the constituent's rank $r_{\rm u}$ targeted by the perturbation.
\revise{Perturbing constituents with small rank (high importance) in comparison to those with high ranks (low importance) tends not only to lead to greater rank changes of single constituents but also to more constituents showing such changes ($\grd\gg0$) overall.
Widely independent of the networks' topologies, we observe that the smaller the rank $r_{\rm u}$ (the higher the importance) of the perturbed constituent, the larger are the changes in the total ranking of the constituents.}
This general relationship can be observed with each of the employed centrality concepts, while the precise functional relationship depends on multiple factors such as network size, network topology, type of perturbation, and centrality concept.

\begin{figure}[htbp]
\centering
\includegraphics[width=\linewidth]{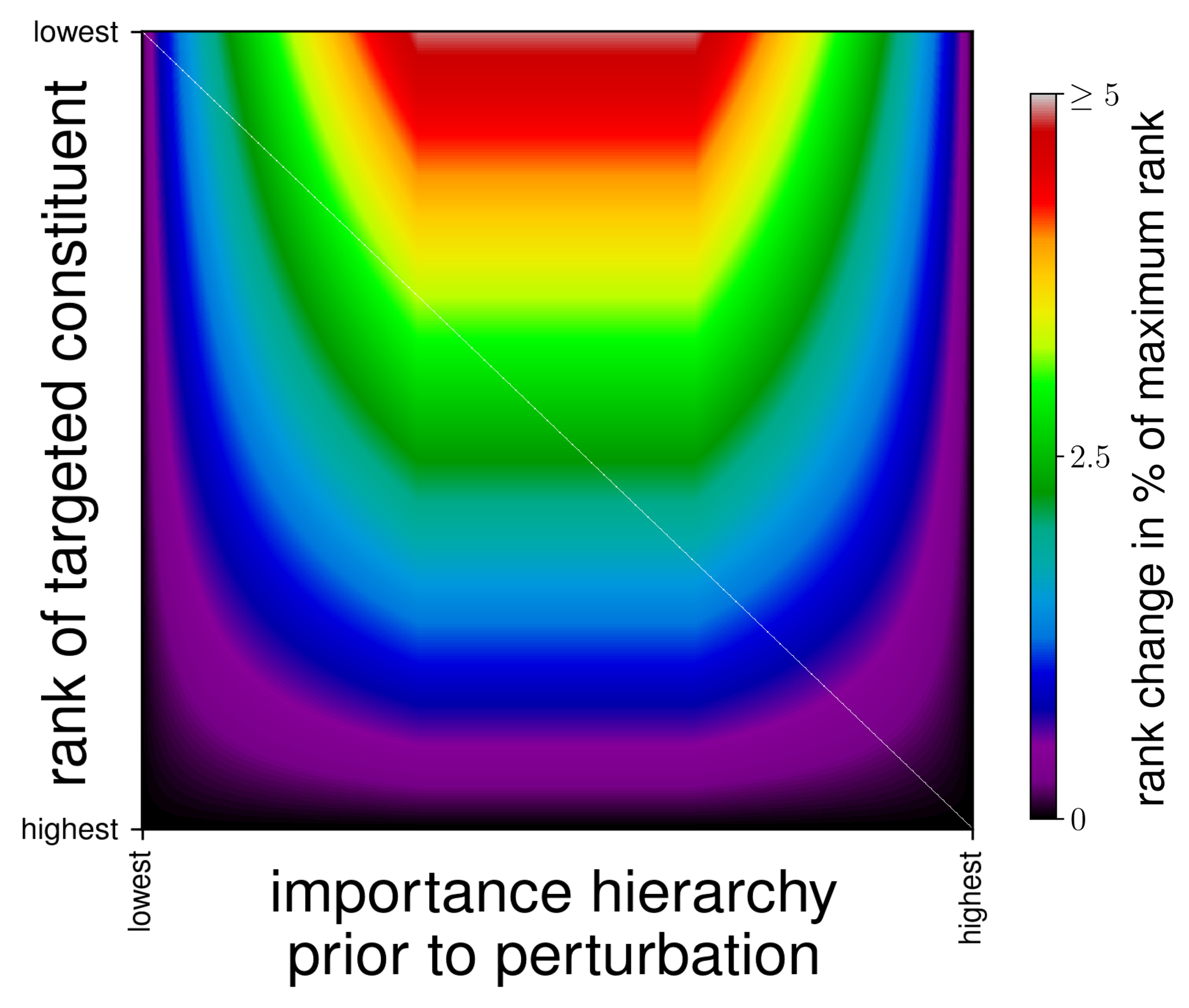}
\caption{Schematic depiction of changes in rank for each constituent (with importance hierarchy prior to perturbation) in dependency of the rank of the targeted constituent.}
\label{fig:fig4}
\end{figure}

The quantitative nature of these relationships regarding perturbation and centrality concept are exemplary depicted in the Appendix (cf. Figures~\ref{fig:a1}-\ref{fig:a4}) for some network models, highlighting not only the overall stronger effect on the ranking when targeting important constituents, but also the fact that the rankings of all the other constituents are affected very distinctly depending on their respective rank prior to perturbation (schematically depicted in Figure \ref{fig:fig4}).
Especially in case of removing vertices, we observe that constituents at either end of the importance hierarchy are affected less by this perturbation than constituents with median rank.
\revise{Our results here are in line with those reported in Sect.~\ref{criteria2} and indicate that particularly those constituents identified as less important prior to perturbation can be deemed potentially superfluous.} 

\revise{
\subsection{Concluding Remarks}
It is to conclude that while aspects such as the path-structure, degree-correlations, robustness, and stability are, largely and on average, left unaltered by the employed perturbations, we do observe high dependency of changes of metrics $\overline{\delta \mu}$ regarding the rank of the constituent targeted by the perturbation. 
This points towards a possible intrinsic existence of potentially superfluous constituents in the networks investigated here. 
Observing changes in the rankings of constituents (as determined with different centrality concepts) consolidates these findings. 
Especially targeting constituents with low rank (high importance) in comparison to targeting constituents with high rank (low importance) led to greater changes in these rankings and revealed dependencies regarding the network topology~\cite{Saavedra2011}.
This shows that --~arguably contrary to expectation~-- less dense networks and also less regular networks (following no trivial geometric arrangements such as a ring or a lattice) can contain more potentially superfluous constituents than for example very dense and even complete network.
For the investigated networks, we can conclude that the three criteria point toward constituents of tendentially low (but not least) importance, to be potentially superfluous.  
}

\section{Discussion}
\revise{We here proposed a perturbation-based method in order to tackle the extensive problem of identifying potentially superfluous network constituents.}
We formulated the premise that the instant absence or additional presence of a \revise{potentially} superfluous network constituent should lead to negligible changes in network characteristics only, that not trivially depend on even the smallest change in network size. 
Making use of minuscule and elemental perturbations, targeting single constituents directly, we investigated whether such perturbations lead to changes \revise{of} global as well as local \revise{metrics} that describe the investigated networks rather comprehensively. 
The less changes we observe for the \revise{metrics} when perturbing the respective network constituent the more this is an indication for this constituent to be \revise{of potentially} superfluous nature.
\revise{We formulated three criteria, which can provide important information when it comes to identifying potentially superfluous constituents.}

It is generally to be expected that, independent of the investigated real-world system, certain network topologies may contain superfluous constituents, simply due to their structural makeup.
Following this line of thinking, it is to be expected that the sheer size and density of complete networks should provide great possibility for the existence of such \revise{potentially} superfluous constituents. 
Likewise, it is easy to understand that certain regular structures, e.g., a ring \revise{or a lattice} with a large amount of nearest- and next-nearest-neighbors connections, are more likely to contain \revise{potentially} superfluous constituents than a ring/lattice with only nearest-neighbors connections.

We could confirm that --~even though generally shown to have small influence on local as well as global characteristics \cite{Holme2002,Platig2013,lekha2020,rings2022}~-- the here investigated effects of employed perturbations indeed \revise{largely differ for different network topologies as revealed with the three criteria.} 
While we could show that size and edge density affect \revise{the values of global network metrics}, the here employed minuscule perturbations, on average, led to negligible changes \revise{of these values}.
On the other hand, regular structures, whether it be a ring/lattice or a complete network, were especially prone to be influenced in their \revise{local characteristics}.

\revise{Although} in almost regular networks an importance hierarchy is dominated by edge weights, said importance hierarchy in less regular networks might be influenced by their distinct topological makeup (small-world or scale-free networks).
This makes it rather hard to get an intuitive feeling about the existence of \revise{potentially} superfluous constituents in these complex network topologies.
However, contrary to expectation, our perturbation-based approach \revise{points} to far less \revise{potentially} superfluous constituents in complete and regular networks than in more complex topologies such as small-world and scale-free networks. 

Nevertheless, independent of the networks' topology, if the perturbation targeted a more important constituent, the changes in \revise{local network characteristics} (distribution of centrality values and centrality-based rankings) were also larger, \revise{in comparison to targeting a less important constituent.}
This also shows that a priori knowledge about the importance hierarchy of the networks' constituents might not only be highly useful but in some cases even necessary to end up with a satisfactory and somewhat accurate representation \revise{of a real-world complex system.}
In addition, and especially in those cases for which a priori knowledge about the network's actual structure is either not accessible or very limited, our perturbation-based approach can aid in identifying \revise{potentially} superfluous and likewise indispensable network constituents.

\revise{Future studies should focus on employing the presented approach to identify potentially superfluous constituents either in networks with built-in superfluous constituents or in networks constructed from empirical observations prone to have superfluous constituents.}
Further investigations considering \revise{scenarios from real-world issues} (like noise contamination and other measurement errors~\cite{Martin2019}) might aid in a more accurate modeling of real complex (dynamical) systems. 
This could mean taking into account not necessarily the exact cloning of network constituents but a combination of cloning and perturbations regarding the edge weights of cloned edges.\\

\begin{acknowledgments}
The authors would like to thank Thorsten Rings for interesting discussions and for critical comments on earlier versions of the manuscript.
\end{acknowledgments}

\section*{Data Availability Statement}
The data that support the findings of this study are available from the corresponding author upon reasonable request.

\appendix
\section*{Appendix}

\begin{figure*}[]
\centering
\includegraphics[width=\linewidth]{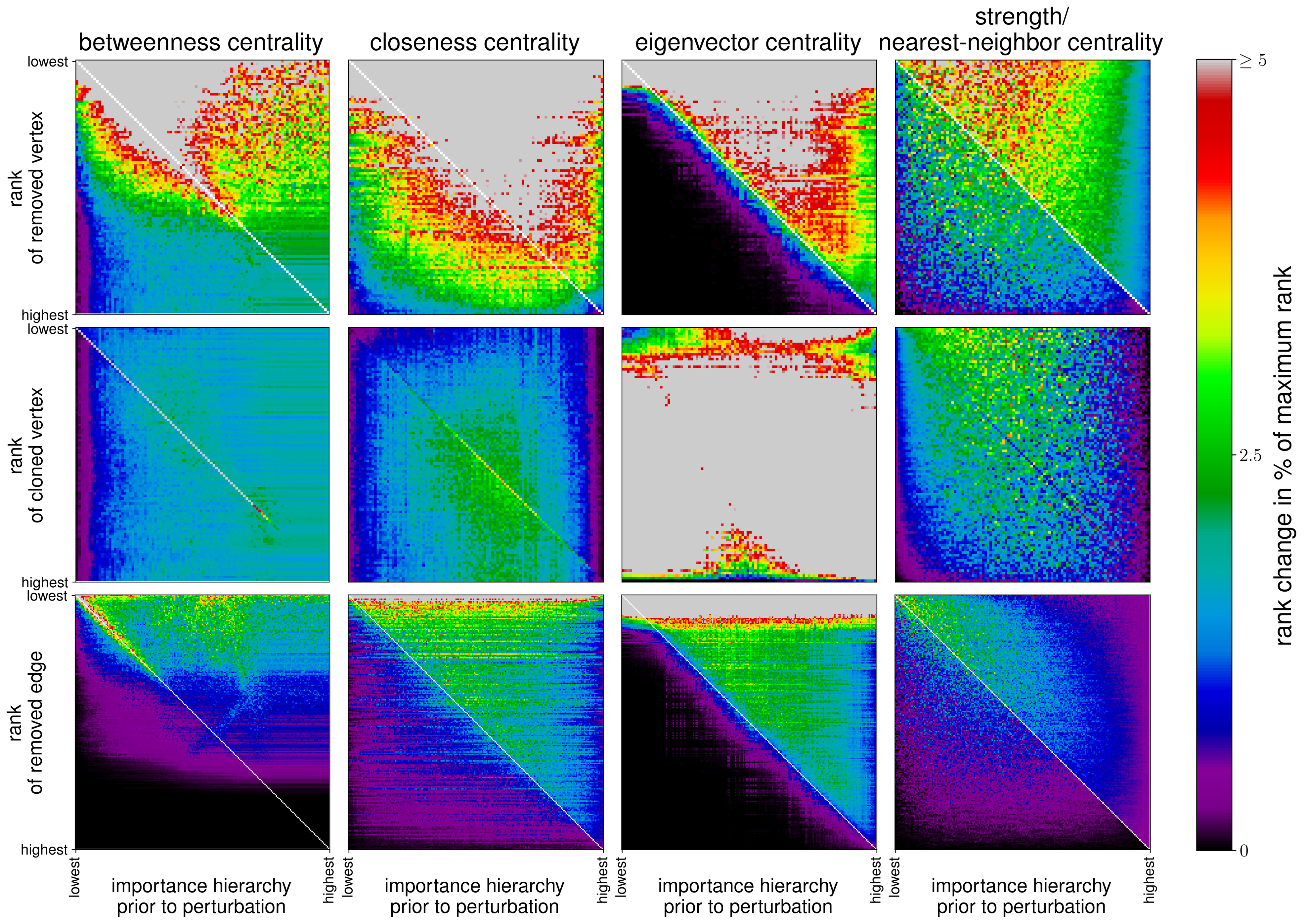}
\caption{Changes in rank for each constituent (with importance hierarchy prior to perturbation) in dependency of the rank of the targeted constituent in a small-world network with $V=100$, $m=4$, $p=0.01$.}
\label{fig:a1}
\end{figure*}

\begin{figure*}[]
\centering
\includegraphics[width=\linewidth]{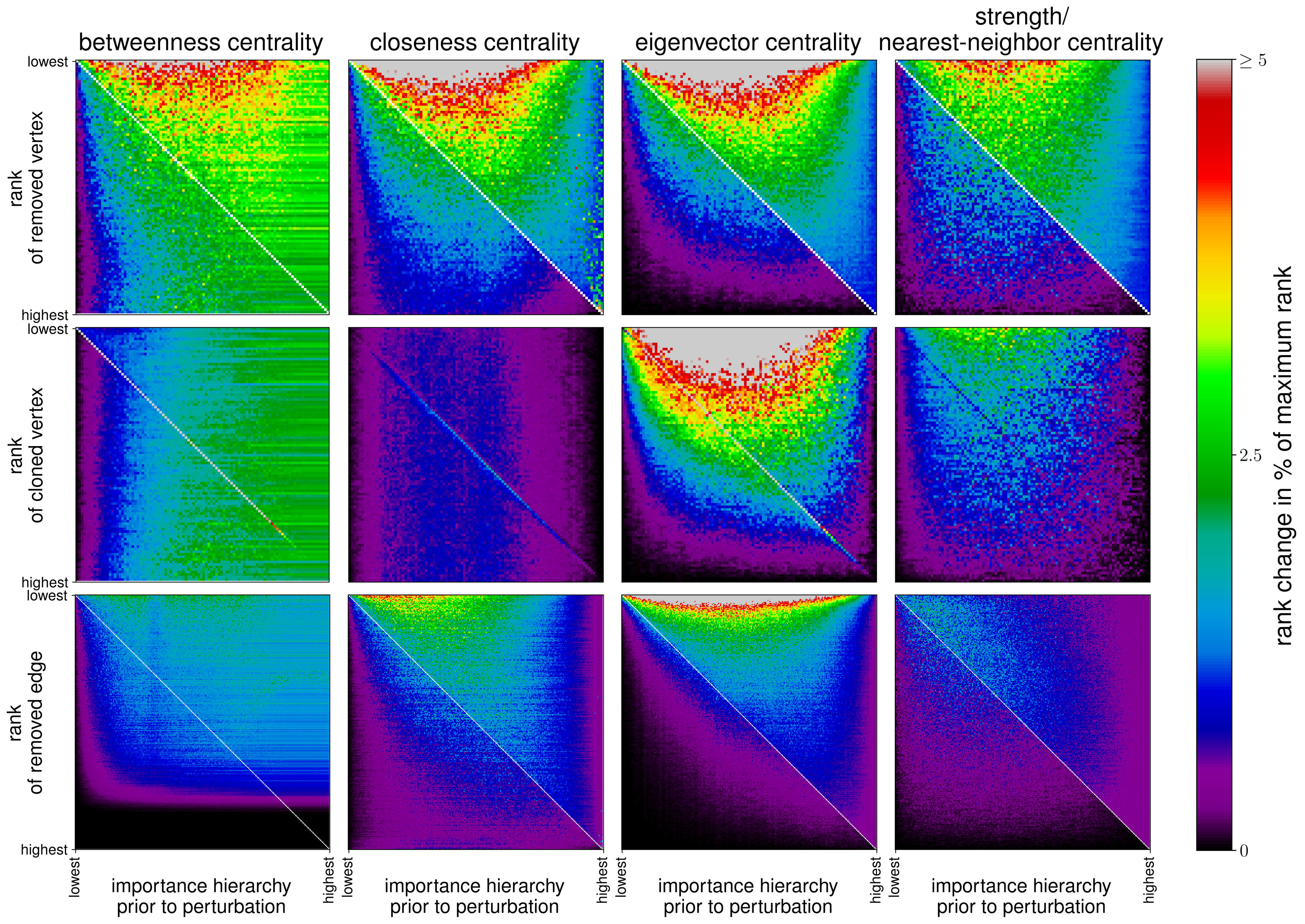}
\caption{Same as Fig.~\ref{fig:a1} but for a random network with $V=100$, $q=0.05$.}
\end{figure*}
\begin{figure*}[]
\centering
\includegraphics[width=\linewidth]{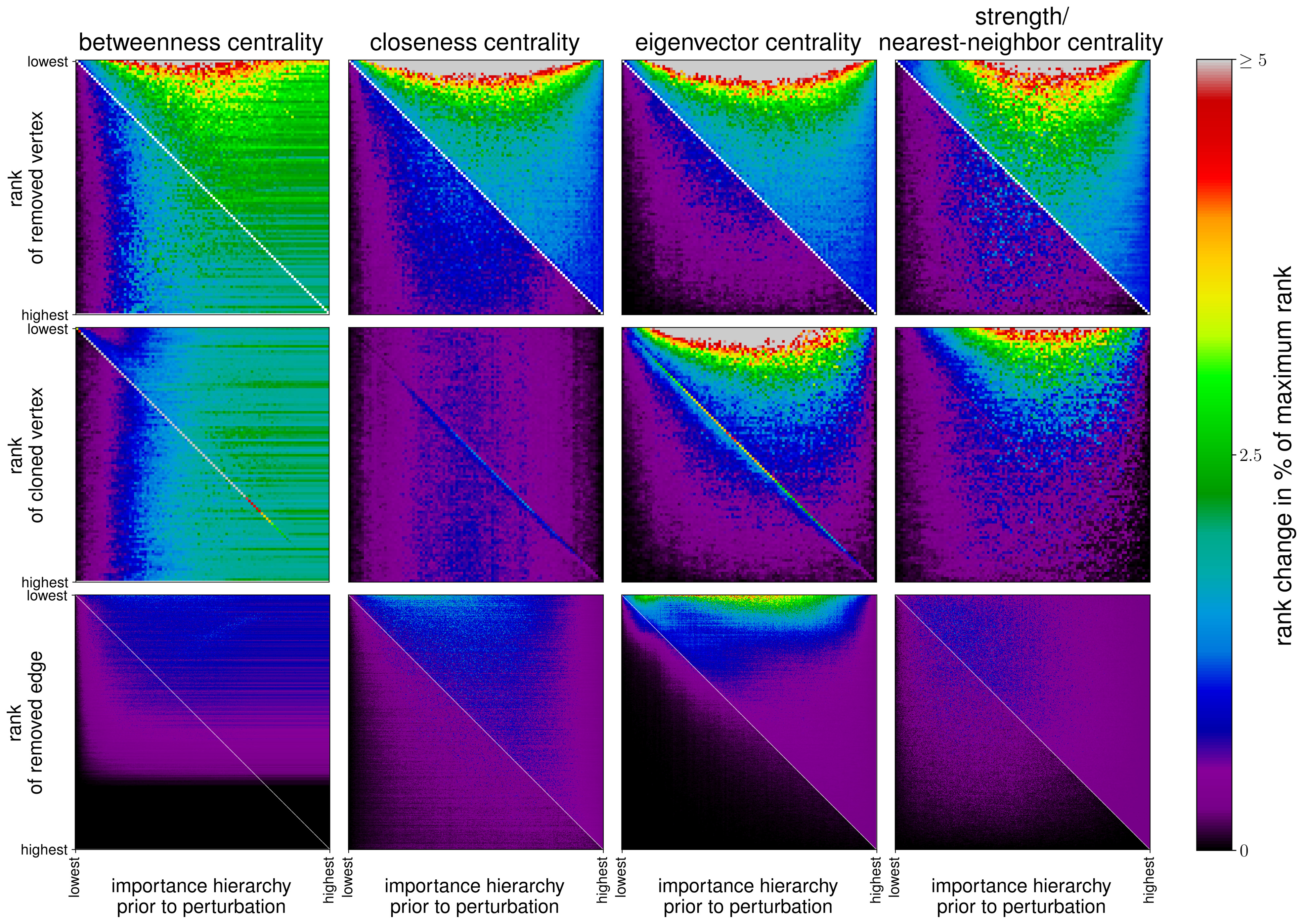}
\caption{Same as Fig.~\ref{fig:a1} but for a scale-free network with $V=100$, $k=4$.}
\end{figure*}
\begin{figure*}[]
\centering
\includegraphics[width=\linewidth]{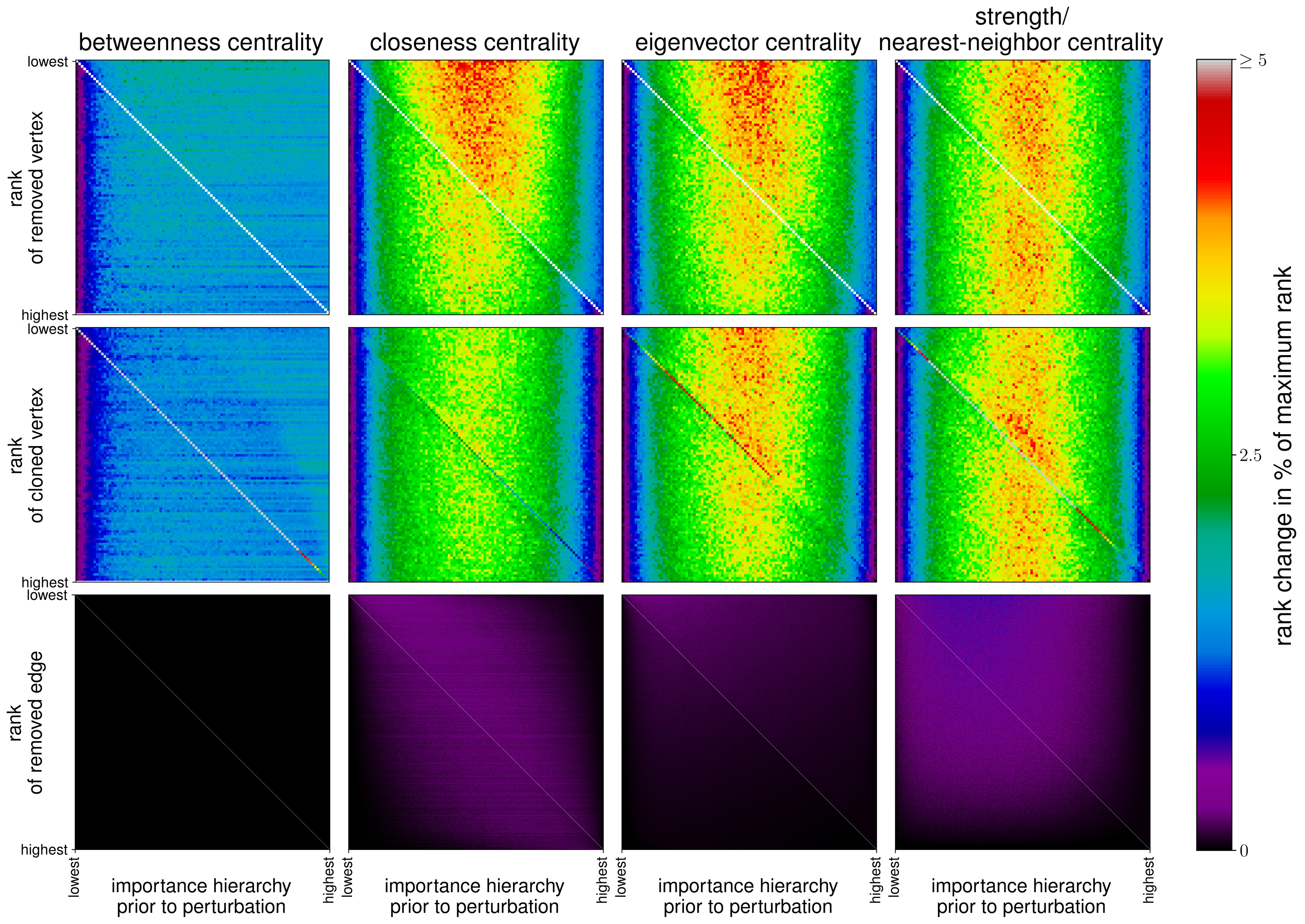}
\caption{Same as Fig.~\ref{fig:a1} but for a complete network with $V=100$.}
\label{fig:a4}
\end{figure*}
%


%

\end{document}